\documentclass[pra,aps,showpacs, amsmath, amssymb, superscriptaddress, showpacs, twocolumn, notitlepage,10pt]{revtex4-1}
\usepackage{color,ulem}
\usepackage{amsfonts}              
\usepackage{amsbsy}
\usepackage{latexsym}
\usepackage{mathrsfs}
\usepackage{bm}
\usepackage[dvipdfmx]{graphicx}
\usepackage{epstopdf}
\usepackage{booktabs}
\usepackage[titletoc,title]{appendix}

\newcommand{\be}{\begin{eqnarray}}
\newcommand{\ee}{\end{eqnarray}}
\def\({\left(}
\def\){\right)}
\def\[{\left[}
\def\]{\right]}

\def\Z{\mathbb{ Z}}

\newcommand{\ve}[1]{ \boldsymbol #1}

\newcommand{\Tr}{\mathrm{Tr}}
\newcommand{\sla}[1]{\rlap{\kern .15em /}#1}

\begin{document}

\title{Construction of Arbitrary Robust One-Qubit Operations Using Planar Geometry}

\author{Tsubasa Ichikawa}
\affiliation{Department of Physics, Gakushuin University, Tokyo 171-8588, Japan}
\author{Jefferson G. Filgueiras}
\affiliation{Fakult\"at Physik, Technische Universit\"at Dortmund, D-44221 Dortmund, Germany}
\author{Masamitsu Bando}
\affiliation{Research Center for Quantum Computing, Interdisciplinary Graduate School of Science
and Engineering, Kinki University, 3-4-1 Kowakae, Higashi-Osaka, Osaka 577-8502, Japan}
\affiliation{Centro Brasileiro de Pesquisas F\'{\i}sicas, Rua Dr. Xavier
Sigaud 150, 22290-180 Rio de Janeiro, Rio de Janeiro, Brazil}
\author{Yasushi Kondo}
\affiliation{Research Center for Quantum Computing, Interdisciplinary Graduate School of Science
and Engineering, Kinki University, 3-4-1 Kowakae, Higashi-Osaka, Osaka 577-8502, Japan}
\affiliation{Department of Physics, Kinki University, 3-4-1 Kowakae, Higashi-Osaka, Osaka
577-8502, Japan}
\author{Mikio Nakahara}
\affiliation{Research Center for Quantum Computing, Interdisciplinary Graduate School of Science
and Engineering, Kinki University, 3-4-1 Kowakae, Higashi-Osaka, Osaka 577-8502, Japan}
\affiliation{Department of Physics, Kinki University, 3-4-1 Kowakae, Higashi-Osaka, Osaka
577-8502, Japan}
\author{Dieter Suter}
\affiliation{Fakult\"at Physik, Technische Universit\"at Dortmund, D-44221 Dortmund, Germany}


\begin{abstract}
We show how to construct an arbitrary robust one-qubit unitary
operation with a  control Hamiltonian of 
$A_x(t) \sigma_x + A_y(t)
 \sigma_y$, where $\sigma_i$ is a Pauli matrix and $A_i(t)$
is piecewise constant.
Our method, based on planar geometry, 
admits a simple and intuitive interpretation. Furthermore,
the total execution time and
the number of elementary gates of the obtained sequence are comparable
to those of the shortest known concatenated composite pulses.  
\end{abstract}
 

 \maketitle

\section{Introduction}
Precise control of quantum systems has been indispensable 
in many research fields in physics. In particular, the nuclear 
magnetic resonance (NMR) community developed the so-called 
composite pulses \cite{Levitt86}, which are pulse sequences designed to
implement operations robust against some types of systematic
errors. Two relevant examples are the
amplitude and off-resonance errors. The first one occurs when 
the amplitude of the driving field deviates from its nominal value.
The off-resonance error happens when the control field is not perfectly resonant 
with the qubit transition.

Composite pulses are also useful in quantum information processing
\cite{NC00, SS08, NO08, Jones11, Merrill12}, since they can be used 
as robust quantum gates whose performance suffers less from the 
systematic errors under consideration  \cite{Jones12}. 
Recently, new sequences have been constructed in the context of 
double quantum dot systems \cite{Wang12, Kestner13} and for qubit 
addressing in ion traps and optical lattices \cite{address}.

In recent works 
\cite{Ichikawa11, Ichikawa13, Bando13, Brown04, Odedra12, Jones13, Cummins03, Knillpulse}, 
several composite pulses were proposed and their properties investigated. 
For example, it was shown that any sequence robust against amplitude errors 
is a geometric quantum gate using Aharonov-Anandan phase \cite{Ichikawa12,kondo}. 
Controlled-NOT and SWAP gates robust against coupling strength error 
were designed for Ising-type interactions \cite{Ichikawa13}. Furthermore, 
propagators with simultaneous robustness to two types of systematic 
errors have been obtained \cite{Ichikawa11, Bando13, Brown04, 
Odedra12, Jones13, Cummins03, Knillpulse}.

Although such accurate operations have been introduced, 
they are intricate in their construction \cite{Low14} or
the target unitary gates are restricted within those rotations with the axes 
in the $xy$-plane \cite{Jones13, 
Uhrig07, Jones13a}.

In this paper, 
we construct composite pulse sequences that implement 
arbitrary SU(2) operations. Our approach is simple enough
to be derived with elementary planar geometry, and can be employed 
to improve the robustness of two sequential gate operations 
collectively.

This paper is organized as follows. In section II, a brief review 
on the subject is presented. The third section contains the main 
result of this paper: the analytical formula of the composite pulse 
to implement general single qubit rotations. This formula is applied 
to the phase and Hadamard gates in section IV to demonstrate that 
our schemes achieves robust gates with high fidelity in both cases.
In section V, we compare the sequences derived here with existing 
composite pulses. The last section is devoted to conclusion and discussions.


\section{Gates, Errors and Composite Pulses}
\label{gecp}
Consider a qubit whose dynamics is generated by a control Hamiltonian
\be 
{\mathcal H} = A_x(t) \sigma_x + A_y(t) \sigma_y,
\label{Hjc}
\ee 
where $A_x(t)$ and $A_y(t)$ are assumed to be piecewise constant and 
$\sigma_i$ is a Pauli matrix. Such Hamiltonian  
is found, for example, in Josephson-junction systems \cite{Makhlin01} and 
electrons floating on liquid helium \cite{e_on_He}, where a dipolar qubit
with controllable level spacing is considered. 
This is quite a general Hamiltonian, which offers universal control of 
a qubit. Furthermore, the Jaynes-Cummings Hamiltonian 
with a classical field can be reduced to Eq.~(\ref{Hjc})
\cite{JC, e_on_He}.

To consider a general robust rotation, we introduce two 
assumptions: i) the absolute calibrations of $A_x$ and 
$A_y$ are not very accurate, and ii) the relative calibration 
between the two control fields is precise. Usually, it is 
easier to calibrate relatively than absolutely, which makes these
assumptions reasonable. With these assumptions, 
it is good to rewrite the control Hamiltonian as 
\be
{\cal H}=A\bm{n}(\phi)\cdot\bm{\sigma}/2,
\label{H}
\ee
where $\bm{n}(\phi)=(\cos\phi,\sin\phi,0)$  and 
$\bm{\sigma}=(\sigma_x,\sigma_y,\sigma_z)$. 
The control parameters $A=(A_x^2+A_y^2)^{1/2}>0$ and 
$\phi=\tan^{-1}A_y/A_x$ are the amplitude 
and phase of the applied control field, respectively. 
Due to the accuracy in the relative calibration between
$A_x$ and $A_y$, only $A$ may have a systematic error.
This is the case for systems controlled by 
resonant pulses with controllable phases, such as radiofrequency,
microwave and laser pulses.

If the parameters $A$ and $\phi$ are time-independent, the Hamiltonian (\ref{H})
leads to the one-qubit unitary gate
\be
(\theta)_\phi=e^{-i{\cal H}t}=\cos(\theta/2)\openone-i
\sin(\theta/2)\bm{n}(\phi)\cdot\bm{\sigma},
\label{R}
\ee
where $\theta=At>0$ and $t$ is the pulse duration. The Planck constant 
$\hbar$ has been set to unity and $\openone$ is the identity operator. 
An unitary gate is trivial if it can be reduced to the identity
operator, i.e., $(\theta)_\phi=\pm\openone$.

As already mentioned, the Hamiltonian (\ref{H}) can be affected 
by systematic errors, due to
imperfections in the control fields. Now let us introduce the two systematic errors we 
are concerned with. One is the amplitude error, which replaces
\be
\theta\rightarrow\theta^\prime=(1+\epsilon)\theta,
\ee
where $|\epsilon|\ll1$ is the magnitude of amplitude error. The other one is 
the off-resonance error, 
an undesired non-zero $\sigma_z$-term in addition to the Hamiltonian (\ref{H}):
\be
{\cal H}\rightarrow A(\bm{n}(\phi)\cdot\bm{\sigma}+f\sigma_z)/2,
\ee
where $|f|\ll1$ denotes the off-resonance error.

Due to these systematic errors, the gate $(\theta)_\phi$ changes into
\be
(\theta)_\phi&\rightarrow&(\theta)_\phi^\prime\nonumber\\
&=&\cos(\theta^\prime/2)\openone-i\sin(\theta^\prime/2)
(\bm{n}(\phi)\cdot\bm{\sigma}+f\sigma_z)\nonumber\\
&\approx&(1-i\epsilon\theta/2)(\theta)_\phi-if\sin(\theta/2)\sigma_z
\ee
to the first order in $\epsilon$ and $f$. The gate $(\theta)_\phi$ is robust against
amplitude (off-resonance) errors if $(\theta)_\phi^\prime$ has no first order term in
$\epsilon$ ($f$). A direct consequence of the above equation is the
robustness of $(2\pi)_{\phi}$ against off-resonance errors for any $\phi$.

Given a target propagator $(\theta)_\phi$ to be implemented, 
the composite pulses are defined as arrays of unitary gates 
which satisfy
\be
(\theta_N)_{\phi_N}^\prime\cdots
(\theta_1)_{\phi_1}^\prime=(\theta)_\phi+{\cal O}(\epsilon^2,\epsilon f,
f^2)
\ee
so that the first order error terms are eliminated.

We first consider a composite pulse robust against amplitude errors.
An arbitrary SU(2) gate can be decomposed
into three rotations with their axes in the $xy$-plane, and whose rotation 
angles are equivalent to the Euler angles.
One possible strategy to make this sequence robust against amplitude errors is to 
replace each elementary gate by a composite pulse insensitive to
such errors. Since available nontrivial composite pulses consist of at least 
three pulses \cite{Bando13}, the resulting robust rotation consists of a 
sequence of nine elementary pulses. However, in principle, just three
elementary operations are necessary to define a robust arbitrary 
unitary: the number of free parameters in three gates is six, whereas the zeroth and
first order perturbation terms with respect to $\epsilon$ must satisfy
three constraints each. Since such a construction is intricate, 
we design an alternative and simple composite pulse for general 
single-qubit gates.

\section{Robust Arbitrary  Rotations}
\label{crar}
As discussed in the last section, the majority of composite pulses are
designed on the assumption that the target has its rotation axis in the
$xy$-plane. This restriction is lifted in this section and we show that
general rotations can be made robust against amplitude and off-resonance 
errors. The construction 
of these robust gates relies on the fact that any SU(2) transformation 
$U$ can be decomposed into the form
\be
U=(\theta_2)_{\phi_1}Z_{\phi_2}(\theta_1)_{\phi_1},
\label{rzr}
\ee
with
\be
Z_\phi=e^{-i\phi\sigma_z/2}.
\ee
Using the identity
$Z_\psi(\theta)_\phi=(\theta)_{\phi+\psi}Z_\psi$ in the right-hand side
of Eq.~(\ref{rzr}), 
we obtain
\be
U=\Theta Z_{\phi_2},
\label{utz}
\ee
where
\be
\Theta=(\theta_2)_{\phi_1}(\theta_1)_{\phi_1+\phi_2}.
\label{theta}
\ee
$Z_{\phi_2}$ also takes the form of $\Theta$, since it can be implemented by
two $\pi$ pulses in the $xy$-plane (see Eq.~(\ref{zdec})).
Thus, it is sufficient to design composite pulses for $\Theta$ 
in order to make the target gate 
$U$ robust. 

\begin{figure}
\begin{center}
\includegraphics[width=2.5in]{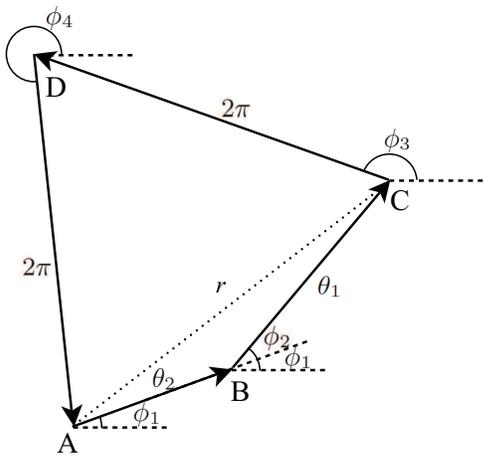}
\caption{Planar geometric interpretation of ${\bm m}=0$. 
The vectors $\protect\overrightarrow{\rm AB}=\theta_2\ve{n}(\phi_1)$, 
$\protect\overrightarrow{\rm BC}=\theta_1\ve{n}(\phi_1+\phi_2)$, 
$\protect\overrightarrow{\rm CD}=2\pi\ve{n}(\phi_3)$ 
and $\protect\overrightarrow{\rm DA}=2\pi\ve{n}(\phi_4)$ are
the summands in Eq.~(\ref{m}). $\theta_1, \theta_2$, and $2\pi$
are the lengths of these vectors, while $\bm{n}(\phi_i)$ indicates
their directions.}
\label{elegeo}
\end{center}
\end{figure}

To design a robust $\Theta$ gate, two trivial gates are added, 
defining the sequence \cite{Brown04}
\be
V=(\theta_2)_{\phi_1}(2\pi)_{\phi_4}(2\pi)_{\phi_3}(\theta_1)_{\phi_1+\phi_2}.
\label{v}
\ee
The robustness of this sequence against amplitude errors is achieved by fixing
$\phi_3$ and $\phi_4$ as functions of the other parameters. Setting 
 $f=0$, we obtain
\be
V^\prime&=&(\theta_2)_{\phi_1}(\epsilon\theta_2)_{\phi_1}(2\pi)_{\phi_4}
(2\pi\epsilon)_{\phi_4}\nonumber\\
&\times&(2\pi)_{\phi_3}(2\pi\epsilon)_{\phi_3}(\epsilon\theta_1)_{\phi_1+\phi_2}
(\theta_1)_{\phi_1+\phi_2}.
\label{ve}
\ee
Since $(2\pi)_\phi$ is the identity operator, this relation reduces to
\be
V^\prime=(\theta_2)_{\phi_1}
W(\theta_1)_{\phi_1+\phi_2},
\ee
where
\be
W=(\epsilon\theta_2)_{\phi_1}(2\pi\epsilon)_{\phi_4}(2\pi\epsilon)_{\phi_3}
(\epsilon\theta_1)_{\phi_1+\phi_2},
\ee
is the total error term for $V$.
To first order in $\epsilon$, $W$ is given by 
\be
W=\openone-i\epsilon\bm{m}\cdot\bm{\sigma}+{\cal O}(\epsilon^2),
\ee
and the error vector {$\bm{m}$} is 
\be
\bm{m}=\theta_1\bm{n}(\phi_1+\phi_2)
+2\pi\bm{n}(\phi_3)+2\pi\bm{n}(\phi_4)+\theta_2\bm{n}(\phi_1).
\nonumber\\
\label{m}
\ee
Thus, $\bm{m}=0$ implies $W=\openone+{\cal O}(\epsilon^2)$,
showing the error cancellation. This condition implies that the four
vectors on the right hand side of Eq.~(\ref{m}) form a quadrilateral 
(See Fig. \ref{elegeo}).

Given a target $U$, the solution of $\ve{m}$ $=$ $0$ can be obtained
from planar geometry (see Appendix) as
\be
\phi_3&=&\pi+\phi_1+\phi_2-\arcsin\(\frac{\theta_2\sin\phi_2}{r}\)-
\arccos\(\frac{r}{4\pi}\),\nonumber\\
\phi_4&=&\pi+\phi_3-\arccos\(1-\frac{r^2}{8\pi^2}\),
\label{p34}
\ee
where
\be
r={\rm AC}=\sqrt{\theta_1^2+\theta_2^2+2\theta_1\theta_2\cos\phi_2}.
\label{defr}
\ee
This is a generalization of the graphical method proposed in \cite{Odedra12, Jones13}.

Once a composite pulse robust against amplitude errors has been designed, 
simultaneous robustness to amplitude and off-resonance errors can be 
achieved if each of the elementary gates
is replaced by a CORPSE pulse\cite{Ichikawa11, Bando13,
Cummins03}. This procedure is called nesting. The CORPSE pulse for the
target $(\theta)_{\phi}$ is the sequence
$(\theta_3)_{\phi_3}(\theta_2)_{\phi_2}(\theta_1)_{\phi_1}$ with
\be
\theta_1&=&2n_1\pi+\theta/2-k,
\qquad
\phi_1=\phi,\nonumber\\
\theta_2&=&2n_2\pi-2k,
\qquad
\phi_2=\pi-\phi,\nonumber\\
\theta_3&=&2n_3\pi+\theta/2-k,
\qquad
\phi_3=\phi,
\label{corpse}
\ee
where $k=\arcsin[\sin(\theta/2)/2]$ and $n_i\in\Z$ $(i=1,2,3)$ 
satisfy a constraint $n_1-n_2+n_3=0$.
The nesting is performed only for the constituent elementary gates that are not
robust against off-resonance errors: $(2\pi)$ pulses need not to be replaced, since they
already have such robustness. Thus, the sequence (\ref{v}) is nested if
$(\theta_1)_{\phi_1 + \phi_2}$ and $(\theta_2)_{\phi_1}$ are replaced by
CORPSE pulses. The resultant nested composite pulse for $\Theta$
consists of eight elementary gates.

\begin{figure}
\begin{center}
\includegraphics[width=2.6in]{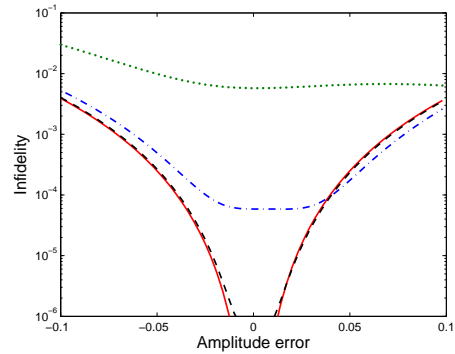}
\caption{(Color online) Infidelity for the composite $Z_\pi$ gate, robust against amplitude errors, 
 as a function of the amplitude error $\epsilon$, for a fixed 
 off-resonance error $f$. The different fixed off-resonant errors are given 
 by $f = 0$ (solid, red), $f = 0.001$ (dashed, black), $f = 0.01$ (dash-dot, blue) 
 and $f = 0.1$ (dotted, green).}
\label{phase}
\end{center}
\end{figure}

\section{Examples}
\label{exa}
In this section, we present two examples of the composite pulse (\ref{v}): 
the $Z_{\phi}$ and the Hadamard gates. In the subsections A and B we 
consider only amplitude errors. The obtained propagators are nested in subsection C
to obtain simultaneous robustness against both amplitude and off-resonance errors.
The robustness against both types of errors is numerically
demonstrated for all examples. The robust $\Theta$ and $Z_{\phi}$ gates
can implement any general robust rotation according to Eq.~(\ref{utz}).

The robustness is calculated in terms of the infidelity between the target
operation $\Theta$ and the implemented composite pulse $V'$:
\be
{\cal I}= 1 - \frac{1}{2}|\Tr(\Theta^\dag V^\prime)|.
\ee
The robustness of the composite pulses will be shown below,
observed through a small gate infidelity up to first order
in amplitude and off-resonant errors.

\begin{figure*}
\begin{center}
(a)\includegraphics[width=2.4in]{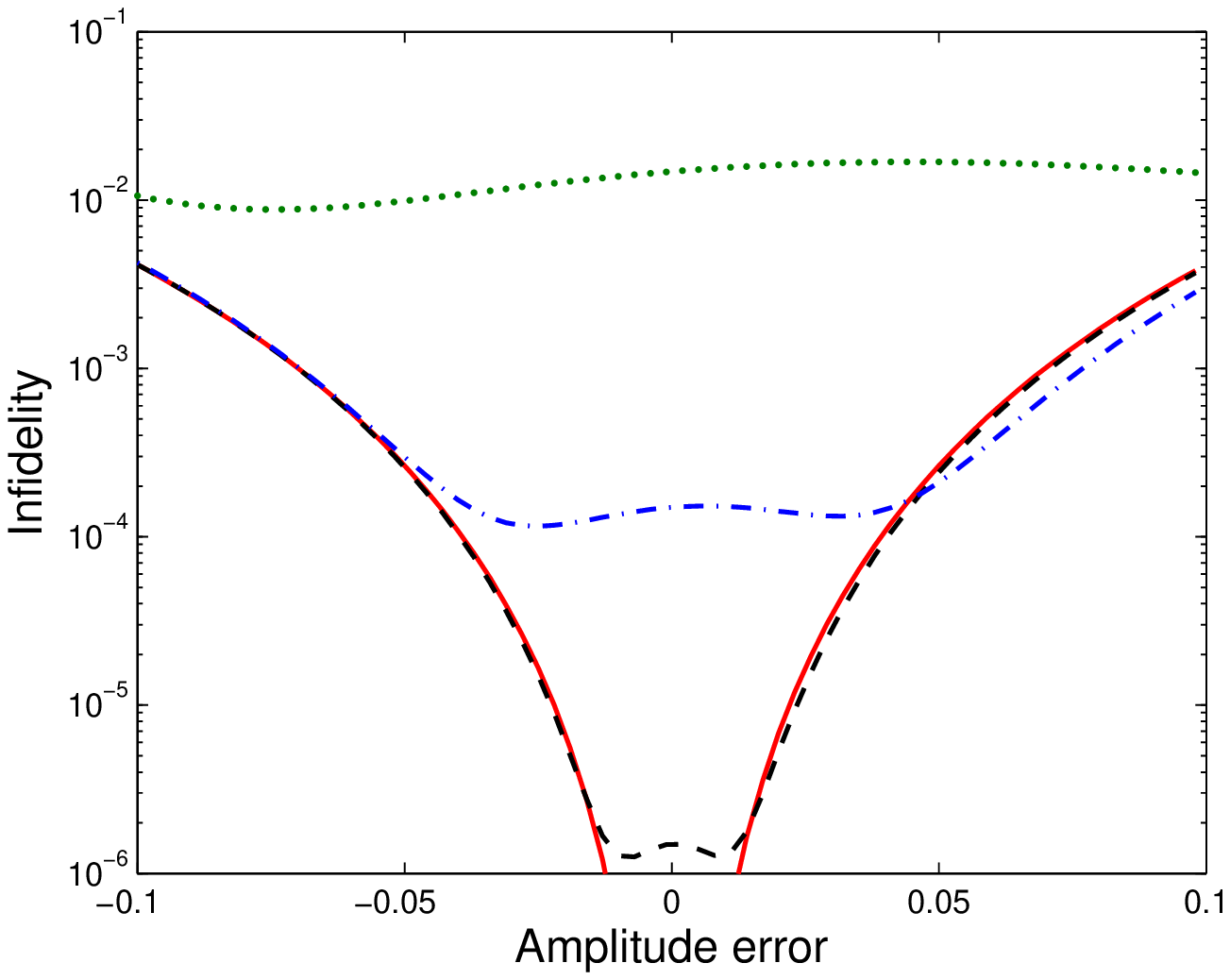}
(b)\includegraphics[width=2.4in]{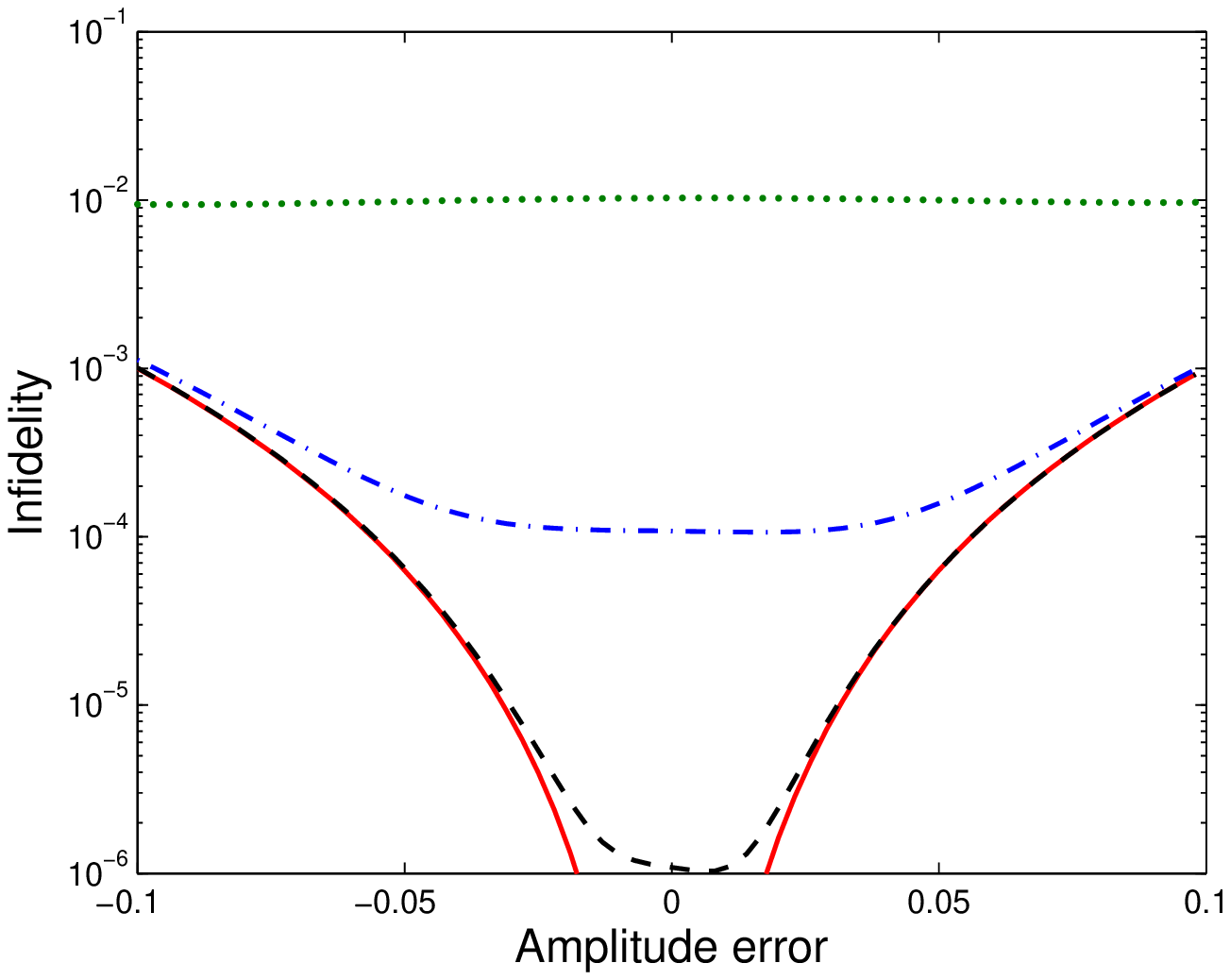}
\caption{(Color online) Infidelity for composite Hadamard gates
 as a function of the amplitude error $\epsilon$, for a fixed 
 off-resonance error $f$. (a) $\Theta = U_H$ based on the decomposition (\ref{h1}).
 (b) $U_{\rm H}$ 
  based on the decomposition (\ref{h2}).
 The different fixed off-resonant errors are given 
 by $f = 0$ (solid, red), $f = 0.001$ (dashed, black), $f = 0.01$ (dash-dot, blue) 
 and $f = 0.1$ (dotted, green).
Note that these Hadamard gates are robust only against amplitude errors.}
\label{hadamard3}
\end{center}
\end{figure*}

\subsection{$Z_\phi$ gate}
\label{h}
First, a robust $Z_{\phi}$ gate is constructed, employing the decomposition 
\be
Z_\phi=(\pi)_0(\pi)_{-\phi/2},
\label{zdec}
\ee
where the unphysical global phase has been ignored. 
Comparing the above expression with Eq.~(\ref{theta}),
$\Theta$ reads
\be
\theta_1=\theta_2=\pi,
\qquad
\phi_1=0,
\qquad
\phi_2=-\phi/2.
\ee
Substituting them into (\ref{p34}) and (\ref{defr}), we obtain
\be
\phi_3&=&\pi-\frac{\phi}{2}-\arcsin\(\frac{-\sin(\phi/2)}{2|\cos(\phi/4)|}\)-
\arccos\frac{|\cos(\phi/4)|}{2},\nonumber\\
\phi_4&=&\pi+\phi_3-\arccos\(1-\frac{\cos^2(\phi/4)}{2}\).
\label{rz}
\ee
Figure~\ref{phase} shows the infidelity between the target gate
$\Theta=Z_\phi$ and the corresponding composite pulse $V^\prime$ with
imperfect pulses.

\subsection{Hadamard gate}
\label{hh}
The composite pulses for the Hadamard gate are obtained for two different decompositions: 
\be
U_{\rm H}&=&(\pi/2)_{3\pi/2}(\pi)_0\label{h1}\\
&=&(\pi/4)_{3\pi/2}(\pi)_0(\pi/4)_{-3\pi/2}.\label{h2}
\ee
The latter decomposition (\ref{h2}) admits a symmetry with
respect to the amplitude of the elementary
pulse. Again, global phases are ignored.

The first decomposition (\ref{h1}) is directly related to $\Theta$ up to a global phase:
\be
\theta_1=\pi,
\quad
\theta_2=\pi/2,
\quad
\phi_1=3\pi/2,
\quad
\phi_2=-3\pi/2,
\ee
which leads to
\be
\phi_3&=&\pi-\arcsin(1/\sqrt{5})-\arccos(\sqrt{5}/8)\approx1.39,\nonumber\\
\phi_4&=&\pi+\phi_3-\arccos(27/32)\approx3.97.
\label{hc2}
\ee
The infidelity is plotted in Fig.~\ref{hadamard3} (b).

To make a composite pulse with the decomposition (\ref{h2}), we set
\be
V=(\pi/4)_{3\pi/2}(2\pi)_{\phi_2}(\pi)_0(2\pi)_{\phi_1}(\pi/4)_{-3\pi/2}
\label{symV}
\ee
and evaluate the error vector $\ve m$. 
Since $(\theta)_\phi(\pi)_0=(\pi)_0(\theta)_{-\phi}$,
we obtain
\be
V^\prime=(\pi/4)_{3\pi/2}(\pi)_0W(\pi/4)_{-3\pi/2}
\ee
with
\be
W=(\epsilon\pi/4)_{-3\pi/2}(2\epsilon\pi)_{-\phi_2}
(\epsilon\pi)_{0}(2\epsilon\pi)_{\phi_1}(\epsilon\pi/4)_{-3\pi/2}.
\nonumber\\
\ee
Expanding $W$ with respect to $\epsilon$, we find
\be
{\ve m}=\pi\begin{pmatrix}
  2\cos\phi_1+2\cos\phi_2+1        \\
  2\sin\phi_1-2\sin\phi_2+1/2       \\
 0
\end{pmatrix}.
\ee
The solution of ${\ve m}=0$ is
\be
(\phi_1, \phi_2)=(\alpha, \beta),
\label{para}
\ee
where
\be
\alpha&=&\arccos((-10-\sqrt{295})/40)\approx2.32,\nonumber\\
\beta&=&\arccos((-10+\sqrt{295})/40)\approx1.39.
\label{hc3}
\ee
The infidelity is given in Fig.~\ref{hadamard3} (b). 
The infidelity profiles (a) and (b) are similar to those 
of the SK1 sequence and the symmetric-BB1 sequence, respectively
\cite{Ichikawa11}. 
These similarities can be understood by the common
feature of  these composite pulses: they are constructed by inserting
several $\pi$ or $2\pi$ rotations whose phases are chosen so that  the
resulting gate is robust against amplitude errors.


\begin{figure*}
\begin{center}
(a)\includegraphics[width=2.4in]{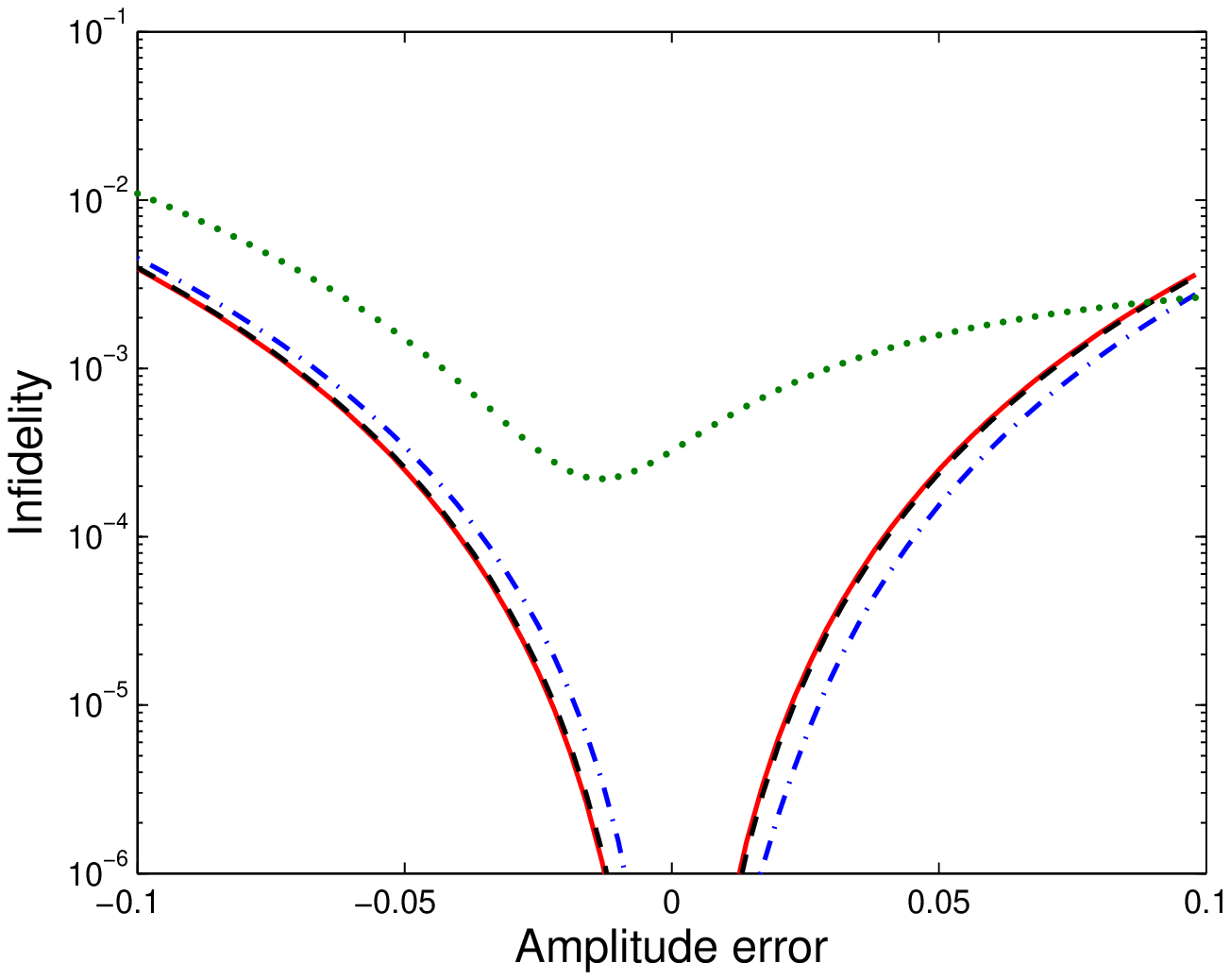}
(b)\includegraphics[width=2.4in]{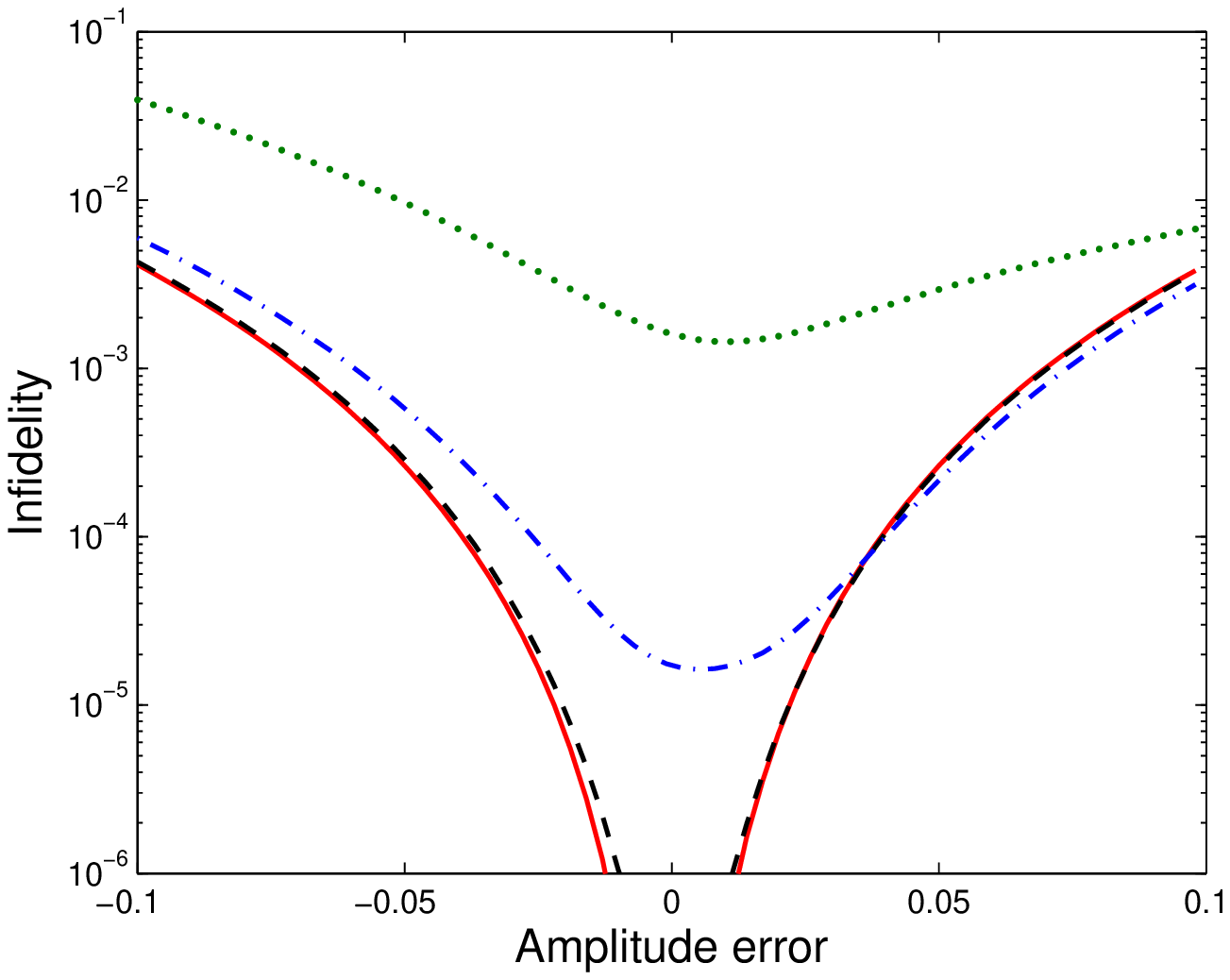}
(c)\includegraphics[width=2.4in]{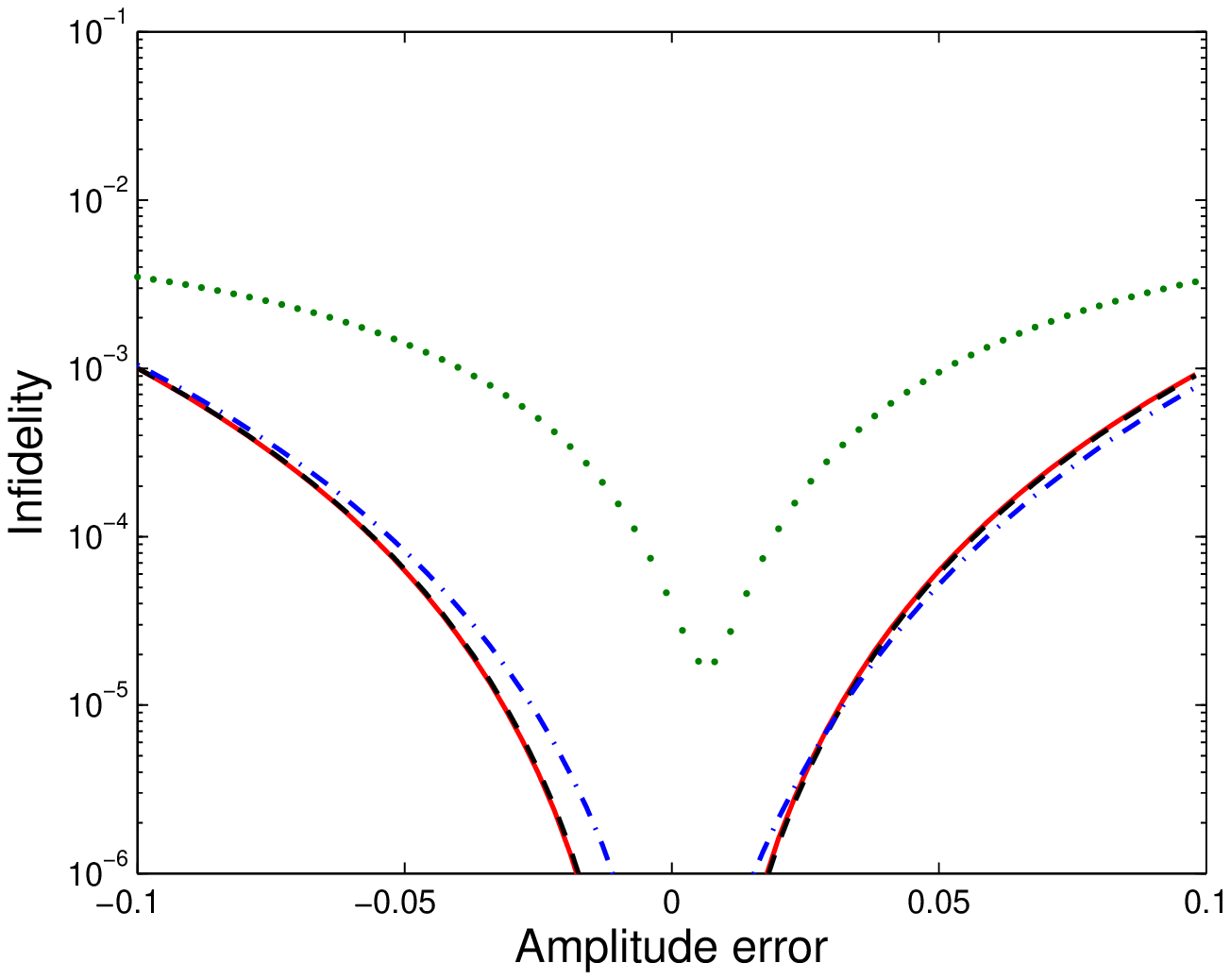}
\caption{(Color online) 
Infidelity for the nested composite pulses, which have simultaneous robustness 
against amplitude and off-resonance errors,
as a function of the amplitude error ($\epsilon$), for a fixed 
ORE ($f$). (a) $Z_\pi$ gate. (b) $\Theta = U_H$ based on 
the decomposition (\ref{h1}). 
(c) $U_H$ 
  based on the decomposition (\ref{h2}).
The different fixed off-resonant errors are given 
by $f = 0$ (solid, red), $f = 0.001$ (dashed, black), $f = 0.01$ (dash-dot, blue) 
and $f = 0.1$ (dotted, green).}
\label{corg}
\end{center}
\end{figure*}

\subsection{Nesting and Simultaneous Robustness}
\label{hs}

Now, the obtained composite pulses are nested so that the resulting gate
is simultaneously robust against amplitude and off-resonance errors. According to the
prescription in Sec.~III, it is sufficient to replace the pulses in
Eqs. (\ref{zdec}), (\ref{h1}), and (\ref{h2}) with CORPSE pulses. For
example, a $(\pi)_0$ pulse is replaced by
\be
(2n_3\pi+\pi/3)_0(2n_2\pi-\pi/3)_\pi(2n_1\pi+\pi/3)_0.
\ee
Figure~\ref{corg} shows the infidelity plots of  the nested sequences.
The infidelity is clearly reduced, even when $\epsilon$ and $f$ are 
both nonzero. The same feature has been observed for other nested 
pulses designed \cite{Ichikawa11, Bando13}.

\subsection{Comparison with other composite pulses}
\label{cocp}
In order to compare different composite pulses, two important criteria
are the number of elementary gates and the operation time cost of the sequence.
The latter is defined by the equation
\begin{eqnarray}
\label{tc}
T = \sum_{i}\frac{\vert\theta_i\vert}{\pi},
\end{eqnarray}
where $\theta_i$ is the flip angle of the $i$-th elementary pulse.
The composite pulse given in Eq.~(\ref{v}) has an operation time cost of
$T_{\rm AE}=4 + (\theta_1 + \theta_2)/\pi$. When nested with CORPSE 
pulses for simultaneous robustness, it is given by 
$T_{\rm nes}=12 + \[\theta_1 + \theta_2 -4(k_1 + k_2)\]/\pi$, where 
$k_i = \arcsin[\sin(\theta_{i}/2)/2]$.

\begin{table}[t]
 \caption{Number of pulses, $N$, operation time
 cost $T$ and robustness of Hadamard and $Z_{\phi}$ gates,
 for different composite pulses.  
 The last column indicates to which error type the sequence is 
 robust, amplitude errors (AE) or off-resonance errors (ORE).}
 \label{tcn}
 \centering
  \begin{tabular}{lccccc}  \hline\hline
   \toprule
   Composite pulse & \multicolumn{2}{c}{Hadamard} & \multicolumn{2}{c}{$Z_\phi$} 
& robustness\\ 
                          & $N$   & $T$       & $N$    & $T$    & \\ \hline
   \midrule
   Eq.~(\ref{v})          & $4$   & $5.5$     & $4$    & $6$    &   AE \\
   Eq.~(\ref{symV})       & $5$   & $5.5$     & --     & --     &   AE \\
   SCROFULOUS             & $6$   & $5.3$     & $6$    & $6$    &   AE \\
   SK1                    & $6$   & $9.5$     & $6$    & $10$   &   AE \\
   BB1                    & $8$   & $9.5$     & $8$    & $10$   &   AE \\
   nested Eq.~(\ref{v})   & $8$   & $12.4$    & $8$    & $12.7$ &   AE, ORE \\
   nested Eq.~(\ref{symV})& $11$  & $16.3$    & --     & --     &   AE, ORE \\
   reduced CinSK          & $10$  & $16.4$    & $10$   & $16.7$ &   AE, ORE \\
   reduced CinBB          & $12$  & $16.4$    & $12$   & $16.7$ &   AE, ORE \\
   reduced SKinsC         & $12$  & $12.4$    & $12$   & $12.7$ &   AE, ORE \\
   \bottomrule \hline\hline
  \end{tabular}
\end{table}

The time cost to make a rotation robust depends 
on its particular decomposition and on which composite pulse 
is employed \cite{Bando13}. To make a comparison 
to the sequence of Eq.~(\ref{v}), $Z_\phi$ and Hadamard gates 
are decomposed as in Eqs.~(\ref{zdec}) and (\ref{h1}),
and each rotation is replaced by a composite pulse.
As it can be seen in Table~\ref{tcn}, both gates can be implemented 
in a shorter time and/or smaller number of pulses if
Eq.~(\ref{v}) is used, in comparison to other 
existing composite pulses. To design gates robust against amplitude errors, 
SK1 and BB1 sequences \cite{Brown04,Winperis} lead to a longer execution time,
while SCROFULOUS pulses \cite{Cummins03} offer a shorter one. When nested, 
the composite pulse of Eq.~(\ref{v}) has a shorter time 
cost in comparison to reduced CinBB and CinSK sequences, 
and fewer pulses than all three reduced CCCP
(ConCatenated Composite Pulses \cite{Ichikawa11}). The nested symmetric 
Hadamard gate of Eq.~(\ref{symV}) also offers a slightly shorter time cost
than those of CinSK and CinBB sequences.

For a general rotation, according to Eq.~(\ref{utz}), we need to apply 
both $\Theta$ and $Z_\phi$ gates. It implies a time cost 
$T=24 + \[\theta_1 + \theta_2 - 4(k_1 + k_2) + 4\pi/3\]/\pi$
when nested. For suppression of amplitude errors, $T=10 + (\theta_1 + \theta_2)/\pi$.
When applying other sequences, a decomposition 
$U=R_{x}(\alpha_1)R_{y}(\alpha_2)R_{x}(\alpha_3)$ leads to a shorter time cost
with all reduced CCCP, specially with reduced SKinC. For this case, 
$T_{\rm SKinC}=18 + \[\alpha_1 + \alpha_2 + \alpha_3 - 4(k_1 + k_2 + k_3)\]/\pi$.
For reduced CinBB and CinSK, the execution time is of the same order as 
our composite pulse, 
$T=24 + \[\alpha_1 + \alpha_2 + \alpha_3 - 4(k_1 + k_2 + k_3)\]/\pi$.
However, for robustness just to amplitude errors, the BB1 and SK1 pulses imply longer execution
times, while SCROFULOUS attains a shorter one.

\section{Conclusion}
\label{con}

In this paper, we obtained composite pulses
robust against amplitude errors, which implement general one-qubit gates.
Simultaneous robustness against amplitude errors and off-resonance error
can be achieved using the nesting procedure. Our approach is especially 
useful for systems where the virtual $z$-rotation or the phase modulation 
techniques cannot be performed efficiently. Furthermore, the proposed sequences 
can improve the tolerance to these errors of any pair of sequential rotations 
in the $xy$-plane.

Two examples were given to illustrate our method, the
phase $Z_{\phi}$ and Hadamard gates. In both 
cases, an expansion of the high fidelity area has been 
shown numerically as in Figs.~\ref{hadamard3} and~\ref{corg}, demonstrating the
robustness of our composite pulses. A sequence symmetric 
in the amplitudes achieves the better performance at the cost of longer 
execution time. This enhancement is due to the fact that symmetries
increase the robustness of pulse sequences in general \cite{Haeberlen,JonesJMR}.
Moreover, using our method these two gates can be implemented with 
shorter execution times and smaller number of elementary pulses.





\begin{acknowledgments}
TI thanks Yukihiro Ota for valuable discussions.
JGF thanks the Brazilian funding agency CNPq [PDE Grant No. 236749/
2012-9]. MB thanks the PCI-CBPF program, funded by the 
Brazilian funding agency CNPq. YK and MN would like to thank
partial supports of Grants-in-Aid for Scientific Research
from the JSPS (Grant No. 25400422). MN is also grateful
to JSPS for partial support from Grants-in-Aid for
Scientific Research (Grant Nos. 24320008 and 26400422).
\end{acknowledgments}

\appendix*
\section{Derivation of Formula (\ref{p34})}
\label{der}

Here it is shown how to solve the condition $\bm{m} = 0$ to
obtain Eq. (\ref{p34}) with the help of Fig.~\ref{elegeo}. 
The following relation holds at the points C and D; 
\be
\phi_3+\angle{\rm ACB}+\angle{\rm ACD}+(\pi-\phi_1-\phi_2)=2\pi
\label{C}
\ee
and
\be
\phi_4+(\pi-\phi_3)+\angle{\rm CDA}=2\pi,
\label{D}
\ee
respectively. Thus, the problem boils down to evaluating $\angle{\rm ACB}$, $\angle{\rm ACD}$
and $\angle{\rm CDA}$. Using Eq. (\ref{defr}) and $\triangle$ABC in Fig.~1, we obtain
\be
\angle{\rm ACB}=\arcsin\(\frac{\theta_2\sin\phi_2}{r}\).
\ee
We also find from $\triangle{\rm ACD}$ that
\be
\angle{\rm ACD}&=&\arccos\(\frac{r}{4\pi}\),\nonumber\\
\angle{\rm CDA}&=&\arccos\(1-\frac{r^2}{8\pi^2}\).
\ee
Substituting them into Eqs.~(\ref{C}) and (\ref{D}), Eq.~(\ref{p34}) is proved.

This graphical proof implies that in general two $2\pi$ pulses are necessary to
make $\Theta$ robust against amplitude errors. 
The reason is that adding one $2\pi$ pulse
cannot close the triangle when $r>2\pi$, and the error vector $\ve{m}$
does not vanish.

\end{document}